\documentclass[iop]{emulateapj}
\usepackage{natbib,graphicx,amsmath,amsthm,ulem,color}
\usepackage{threeparttable}

\newcommand{\be}{\begin{equation}}
\newcommand{\ee}{\end{equation}}

\newcommand{\dd}{{\partial}}

\newcommand{\icarus}{\rm Icarus}

\def\deg{^\circ}

\renewcommand{\vec}[1]{\mathbf{#1}}

\begin{document}

\title{Stability of the Kepler-11 System and its Origin}
\author{Nikhil Mahajan$^1$, Yanqin Wu$^1$}
\affil{$^1$Department of Astronomy and Astrophysics, University of Toronto, Toronto, ON M5S 3H4, Canada}

\begin{abstract}
A significant fraction of {\it Kepler} systems are closely-packed, largely coplanar and circular. We study the stability of a 6-planet system, Kepler-11, to gain insights on the dynamics and formation history of such systems. Using a technique called `frequency maps' as fast indicators for long-term stability, we explore the stability of Kepler-11 system by analyzing the neighbourhood space around its orbital parameters. Frequency maps provide a visual representation of chaos and stability, and their dependence on orbital parameters. We find that the current system is stable, but lies within a few percent of several dynamically dangerous 2-body mean-motion resonances. Planet eccentricities are restricted below a small value, $\sim 0.04$, for long-term stability, but planet masses can be more than twice their reported values (thus, allowing for the possibility of mass-loss by past photoevaporation). Based on our frequency maps, we speculate on the origin for instability in closely-packed systems. We then proceed to investigate how the system can have been assembled. The stability constraints on Kepler-11 (mainly, eccentricity constraints) suggest that if the system were assembled {\it in-situ}, a dissipation mechanism must have been at work to neutralize eccentricity excitation. On the other hand, if migration was responsible for assembling the planets, there has to be little differential migration among the planets, to avoid them either getting trapped into mean motion resonances, or crashing into each other.
\end{abstract}

\section{Introduction} 
\label{sec:intro}

The {\it Kepler} space mission has successfully detected thousands of extra-solar planetary candidates, using transit photometry.  Among the many stars that are monitored by the telescope, Kepler-11 (a solar-type star) stands out as having six confirmed planets in a tightly packed configuration \citep{LissauerNature}. The six planets, either super-Earth or Neptune in sizes, orbit in roughly the same plane and likely have small eccentricities.  Though remarkable, the Kepler-11 system is not alone: systems like KOI-435, KOI-351 and KOI-2433 have similar or more numbers of detected planets. They may well be typical of a large population of exo-planetary systems -- the probability of 6 planets transiting in front of our line-of-sight being much smaller than just one or a couple planets transiting.  As such, Kepler-11 affords us a unique opportunity to study the formation of these planetary systems, one that should be thoroughly exploited.

In this study, we use numerical simulations to address the following questions: for the observationally determined Kepler-11 parameters, is the system stable? Is the tightly packed Kepler-11 system perched perilously on the cliff of instability? If not, how far is it from the cliff? Would this distance inform us about the past history along which Kepler-11 planets are assembled?

\citet{LissauerNature} conducted the primary photometric analysis of the system and determined planetary properties using the method of transit-timing variations \citep{HolmanMurray,Agol}. Their preferred values for the planet masses and eccentricities are reproduced in Table \ref{table:elements} here. Based on transit probability, they suggested that the mutual inclinations between different orbits are small, possibly of order $1-2\deg$. Moreover, none of the planet pairs are actively engaged in first-order mean-motion resonances (MMRs). But the tight spacing means they are never far away from one. As such, dynamical stability is a serious concern. \citet{LissauerNature} numerically integrated the system over $2.5\times 10^8$ yrs, using their best fit parameters, and found that their most eccentric model (inner most planet $e = 0.05$) is chaotic and unstable. This raises the suspicion that the real system is perched very close to the stability threshold.

A later reanalysis of Kepler-11 by \citep{Lissauer2013}, using data over a longer timespan, produces a different set of planet mass and eccentricity determinations. In particular, the two inner-most planets see their masses reduced by factors of $2-3$, and in connection, their eccentricities much raised. The new masses are now in formal disagreement with the older \citep{LissauerNature} values. It is unclear which set of solution is more likely correct, and it is interesting to investigate what stability has to say about this. 

\citet{Migas} performed a photo-dynamical analysis of Kepler-11, a problem with up to $50$ degrees of freedom. They selected a number of regular solutions, among the large number of possible ones, using a proxy for the Lyapunov exponent, obtained over short numerical integrations ($8000$ yrs). These short-term regular solutions were then integrated for longer periods ($40,000$ yrs) to establish their feasibility. Most of them turn out to be self-disrupting. Even the rare ones\footnote{In particular, model IV in that paper, which is compatible with the \citep{Lissauer2013} solution in planet masses, but not in planet eccentricities.}  that remain stable over the longer timescale appear to be isolated islands embedded among a sea of chaos. \citet{Migas} argued that the chaos is likely a result of 3/4-body MMRs. And the delicate state of their stable solution lead them to suggest that Kepler-11 could only have reached its present state through migration and resonant trapping, a conclusion not shared by our work.

These works highlight two related issues for studying the Kepler-11 system. First, the mass and eccentricities values are not uniquely pinned down. \citet{LXW} demonstrated analytically that there is an inherent degeneracy between mass and eccentricity in interpreting the TTV data. Even when one is able to search a wide range of parameter space, the final mass solution depend on, e.g., what prior one adopts for the orbital eccentricities. For instance, the prior that all possible eccentricity values are flatly distributed in linear space (where most phase space is in high eccentricity), or flatly distributed in logarithmic space (where most phase space is in low eccentricity), lead to different mass determinations. The former assumption lead to more eccentric solutions and lower planet masses \citep[and may have led to the results in][]{Lissauer2013}. The second issue is stability. Stability of the system depends on both planet masses and eccentricities (as well as relative inclinations). Given the large uncertainties involved in these parameters, one may be led to very different conclusions regarding system stabilities.

In order to gain any insight about planet formation, one would also need to employ new techniques when studying system stability. Ideally,
one would like to obtain information about long-term stability by short integrations -- this allows a large parameter space to be covered. Moreover, one would like to have the stability results easily visualized -- this allows a diagnosis for the origin of instability. These two wishes, we argue, may best be fulfilled by the technique of frequency map, first introduced by \citet{Laskar90,Laskar92}. After extensive numerical explorations of the stability landscape, around the \citet{LissauerNature} system parameters, we proceed to infer the most likely formation mechanism for Kepler-11 and similar systems.

\begin{table*}
  \centering
  \caption{The standard system parameters for the Kepler-11 system}
  \label{table:elements}
  \begin{threeparttable}
  \begin{tabular}{ c || c  c  c  c  c  c  c  c }
    \hline \hline
    Planet & Period (days) & $a$ (AU) & $e\sin{\omega}$ &
    $e\cos{\omega}$ & $I$ ($\deg$) & $\Omega$ ($\deg$) & $M$ ($\deg$) & Mass ($M_p/M_* \times 10^6$) \\ 
    \hline
    Kepler-11b & 10.3062  & 0.091110 &  0.0030 & -0.0009   & 88.5 & 0 & 26.1  & 12  \\
    Kepler-11c & 13.0240  & 0.106497 & -0.0026 &  0.0011   & 89.0 & 0 & 169.6 & 36 \\
    Kepler-11d & 22.6823  & 0.154141 & -0.0127 &  0.0148   & 89.3 & 0 & 350.9 & 23  \\
    Kepler-11e & 32.0027  & 0.193972 & -0.0161 &  0.000005 & 88.8 & 0 & 77.2  & 28  \\
    Kepler-11f & 46.6908  & 0.249498 & -0.0119 & -0.0037   & 89.4 & 0 & 23.0  & 10  \\
    Kepler-11g & 118.3812 & 0.463981 &  0      & 0         & 89.8 & 0 & 34.5  & 28  \\
    \hline
  \end{tabular}
  \begin{tablenotes}[para,flushleft]
    {\footnotesize Planet orbital elements and masses, adopted from the all-eccentric model of \citet{LissauerNature}. The stellar mass is set to be $0.95 M_\odot$, and the mass of Kepler-11g is evaluated so that it has a similar density as planets $e$ and $f$. Due to its large separation from the other planets, the exact value of the latter is not dynamically significant.}
  \end{tablenotes}
  \end{threeparttable}
\end{table*}

\section{Numerical Tools}

\subsection{Numerical Integrator}

We numerically integrate the dynamical evolution using the Wisdom-Holman symplectic algorithm \citep{Wisdom1991} in the SWIFTER integration package \footnote{http://boulder.swri.edu/swifter}, an improved version of the original SWIFT package by Levison \& Duncan. The integration results are invalid whenever planets undergo close encounters, but this does not pose issue for our stability study. An additional potential, $U_{GR} = -3(G M_{\star}/c r)^2$, corrects for the first-order post-newtonian effect. For our studies of planet migration (\S \ref{subsec:migration}), we further specify some simple damping formula. In all simulations, a small time-step of $10^{-3}$ years is used, corresponding to $\sim 1/30$ of the orbital period of the inner most planet. The stellar mass is set to be $0.95 M_\odot$ \citep{Lissauer2013}.

We adopt the all-eccentric model of \citet{LissauerNature} as our default set of system parameters (Table \ref{table:elements}).  The longitudes of ascending nodes ($\Omega$) are assumed to be zero for all planets, and the initial mean anomalies are calculated using the transit epochs. Integrated over $250$ Myrs, the system remains stable \citep[confirming][]{LissauerNature}, with little orbital evolution. Eccentricities for the inner two planets, initialized near zero, oscillate to values as large as $0.025$,\footnote{This may suggest that the measured eccentricities are too low -- the system is likely to be observed in its average state rather than at extrema.} while those for the outer planets remain below $0.03$.

As mentioned above, the more recent \citet{Lissauer2013} solution for the Kepler-11 planets differ from that in \citet{LissauerNature}. The orbital eccentricities for planets $b$ \& $c$ are now $\sim 10$ times higher, and, to explain the TTV data, masses of planets $b$ \& $c$ have to be reduced by a factor of $2 - 3$. These changes could substantially affect stability. We observe that, in the \citet{Lissauer2013} solution, the eccentricity vectors, ($e\cos\omega, e\sin\omega$), obtained using three different renditions of TTV data are often orthogonal to each other, indicating that the solutions have landed in very different local minima. It is unclear to us that the final solution, taken to be the algebraic average over these three sets of vectors, also possess small $\chi^2$. Moreover, the currently determined planet masses may be sensitive to the prior adopted for planet eccentricities. So we have decided to adopt the old \citet{LissauerNature} solution but investigate the change in stability as eccentricities and masses are varied.

\subsection{Frequency Maps}

One integration of Kepler-11 for $5$ Gyrs would require $\sim 4000$ hours on a modern desktop computer. This is impractical for extensive explorations. Here, we perform a multitude of short-duration integrations (typically $5$ Myrs), and use the technique of frequency map, first introduced by \citet{Laskar90,Laskar92}, to predict the long-term stability of Kepler-11 system. The frequency map is essentially a Fourier transform of a suitable dynamical variable over the integration time.

The use of the so-called `fast indicators' is essential to dynamical studies, and frequency map is one of these. The other widely used one is the so-called Lyapunov exponents\citep[e.g.][]{Froes,Froes97,Goz}, for which the MEGNO number used in \citet{Migas} is one proxy.  While the Lyapunov exponents characterize the divergence of nearby orbits in physical space, the frequency analysis focuses on the frequency space. There has been much study of the relationship between Lyapunov timescales and macroscopic instability timescales. It is known that some dynamical systems, while exhibiting chaos and diffusion in physical space (short Lyapunov time), are rather regular in frequency space, and, remain stable for an extremely long period of time, the so-called `confined chaos', or `stable chaos', or `Nekhoroshev regime' \citep{Nek,morbi96}. The outer solar system is one such example \citep{MurrayHolman,Ito}, so are many asteroids \citep[e.g.]{Milani,Milani97,Franklin,Guzzo} and Pluto \citep{sussman,Wisdom1991}. \citet{murray97} have investigated this behaviour and argued that the instability timescales are not monotonically related to the Lyapunov exponents.\footnote{In practice, \citet{Winter} showed that the Lyapunov exponent measured using radial excursion, not phase excursion, may be a better stability indicator.} Rather, they depend on the detailed phase space structure, e.g., the presence of KAM barrier, and the prevalence of resonance overlap regions.

The frequency analysis, on the other hand, builds on the fact that nearly integrable systems have quasi-periodic motions. So the Fourier transform of a dynamical variable in such a system will exhibit the same fundamental frequencies, regardless of the integration era, while frequencies for a system on chaotic orbits will exhibit time-varying frequencies -- chaos is related to resonance angles morphing from libration to circulation; crossing the separatrix brings about frequency shifts. The amount of frequency modulation gives one a measure of the size of the chaotic zone \citet{Laskar90}. When tested on the solar system planets, it is found that the motion of Mercury, which is chaotic and undergo significant macroscopic diffusion, exhibit large frequency variations \citep{Laskar92,Lithwick2011}, while those of the outer planets, also chaotic but extremely stable, appears immutable in frequency space \citep{Laskar92}. This motivates us to adopt the frequency analysis to study the long-term stability of Kepler-11 system.

For our problem at hand, every planet has six orbital elements, $\{ a,e,I,\Omega, \omega, \lambda \}$ \citep[notation a la][]{MD00}. Which would be a suitable dynamical variable for frequency map?

Since the Kepler-11 system is not in any low-order MMRs, its dynamics is likely dominated by secular interactions, where planets interact gravitationally as if they are massive wires. One can average over the mean longitudes to obtain the orbit-averaged Hamiltonian, $H$, and the equations of motion are,
\begin{equation}
  \frac{da_i}{dt} = 0\,, \,\,\,\,\,\,\,\,\,\, \frac{dz_i}{dt} = i\frac{\dd H}{\dd
    z_i^*}\, ,\,\,\,\,\,\,\,\,\,\,
  \frac{d\zeta_i}{dt} = i\frac{\dd H}{\dd \zeta_i^*}\, ,
\end{equation}
where the subscript $i=1,...N$ is the planet index, $z \approx e\exp(i\varpi)$ and $\zeta \approx I \exp(i\Omega)$ are the complex eccentricity and inclination variables, $z^*$, $\zeta^*$ their complex conjugates, and the angle $\varpi = \omega + \Omega$. Semi-major axis will not vary in secular interactions and is therefore not a suitable variable to characterize the system, while $z$ and $\zeta$ appear promising.

\begin{figure}
	\centerline{\includegraphics[width=0.49\textwidth,trim=0 0 0 0,clip]{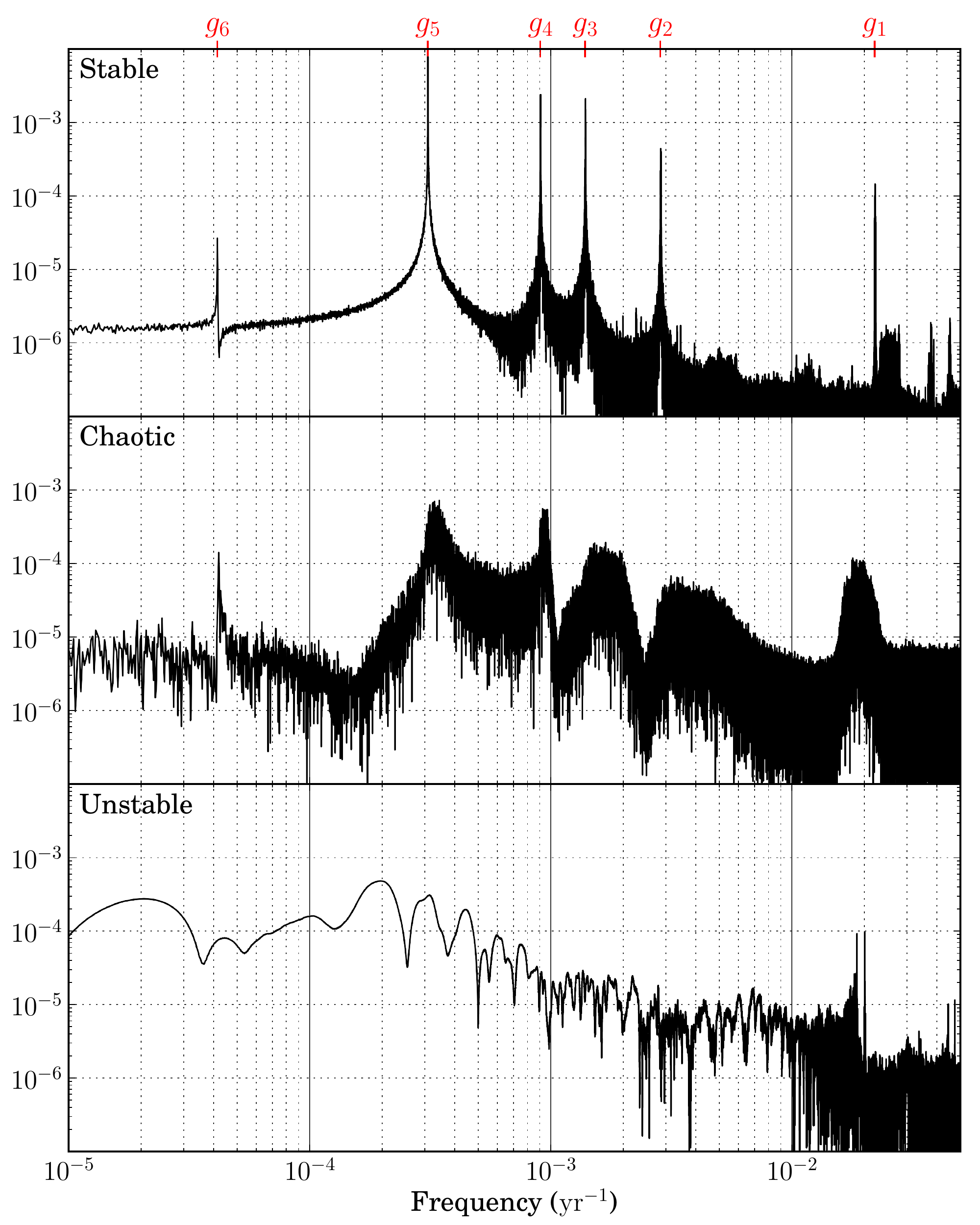}}
	\caption{\footnotesize Fourier Transform of $z$ of Kepler-11c as it is placed at different semi-major axes: observationally determined $a_c$ (stable), $1.03 a_c$ (chaotic) and $0.978 a_c$ (unstable). The Fourier amplitude of the unstable system is several orders of magnitude above the stable and chaotic ones. Labelled on top are the locations of the $6$ linear eigenfrequencies for the eccentricity.}
	\label{fig:frq}
\end{figure}

In cases of small eccentricity and inclination, the secular equations can be reduced to a linear eigenvalue problem and one where $z$ and $\zeta$ motions are decoupled \citep[the Laplace-Lagrangian theory, see][]{MD00}. Frequency analysis of $z_i$ should reveal a set of $N=6$ eigenfrequencies, and that of $\zeta_i$ should give another set ($5$ unique frequencies instead of $6$ as one of the frequencies is zero). The dynamics of such a system is regular, and periodic (top panel of Fig. \ref{fig:frq}).

At larger values of eccentricity and inclination, or when the system lies close to some low-order MMRs, the motion of $z$ and $\zeta$ are no longer regular.  A frequency analysis of $z_i$ (or $\zeta_i$) could show either dominant peaks accompanied by lifted shoulders (middle panel of Fig. \ref{fig:frq}), when the dynamics is mildly nonlinear; or nearly feature-less broad-band (lower panel of Fig. \ref{fig:frq}), when the system is highly chaotic. Close-encounters are destined to occur in such a system.

Since any chaos or instability in the system would affect the frequency behaviour of both $z$ and $\zeta$, analysis of either one is sufficient. We focus on $z$ in this study. \cite{Laskar92} showed that this is a suitable choice even in cases where $z$ is not strictly an action-angle variable.

In our study, we explore how the system stability changes when, e.g., the semi-major axis of planet $c$, is varied over a small range. Instead of plotting Fourier spectrum of every integration, we compress the information into a map (see, e.g., Fig. \ref{fig:sm-a}), where the horizontal axis is the inverse frequency, the vertical axis the parameter that is varied, and the brightness of the pixel represents the Fourier amplitude. Stable regions show up as black backdrop perpetrated by crisp white peaks (location of the linear eigenfrequencies); chaotic or unstable systems are characterized by horizontal striations, indicating that their Fourier power has spilled into a wide band of frequencies. The most unstable systems (ones with close encounters during our short integration span) are picked up as uniformly white stripes, indicating that they have no regularity in the motion whatsoever.

It is interesting to observe that the eigenfrequencies and Fourier amplitudes undergo large swings even in the small range that we explore. We expand on this below.

\section{Stability of the Kepler-11 system -- Parameter Exploration}
\label{sec:stability}

We generate clones of the standard case, integrate them for typically $5$ Myrs, and calculate the frequency map. The clones differ from the standard case in only one parameter. We have chosen to explore how the stability is affected by the semi-major axis, or the eccentricity or the mass of a given planet. These parameters are most revealing about the formation history. We adopt a working definition for stability: a system is considered stable if its Fourier transform exhibits crisp peaks, with amplitudes at least an order of magnitude above
neighbouring bands.

\subsection{Semi-major axis}

\begin{figure*}
	\centering
	\includegraphics[width=0.495\textwidth,trim=0 0 0 0,clip]{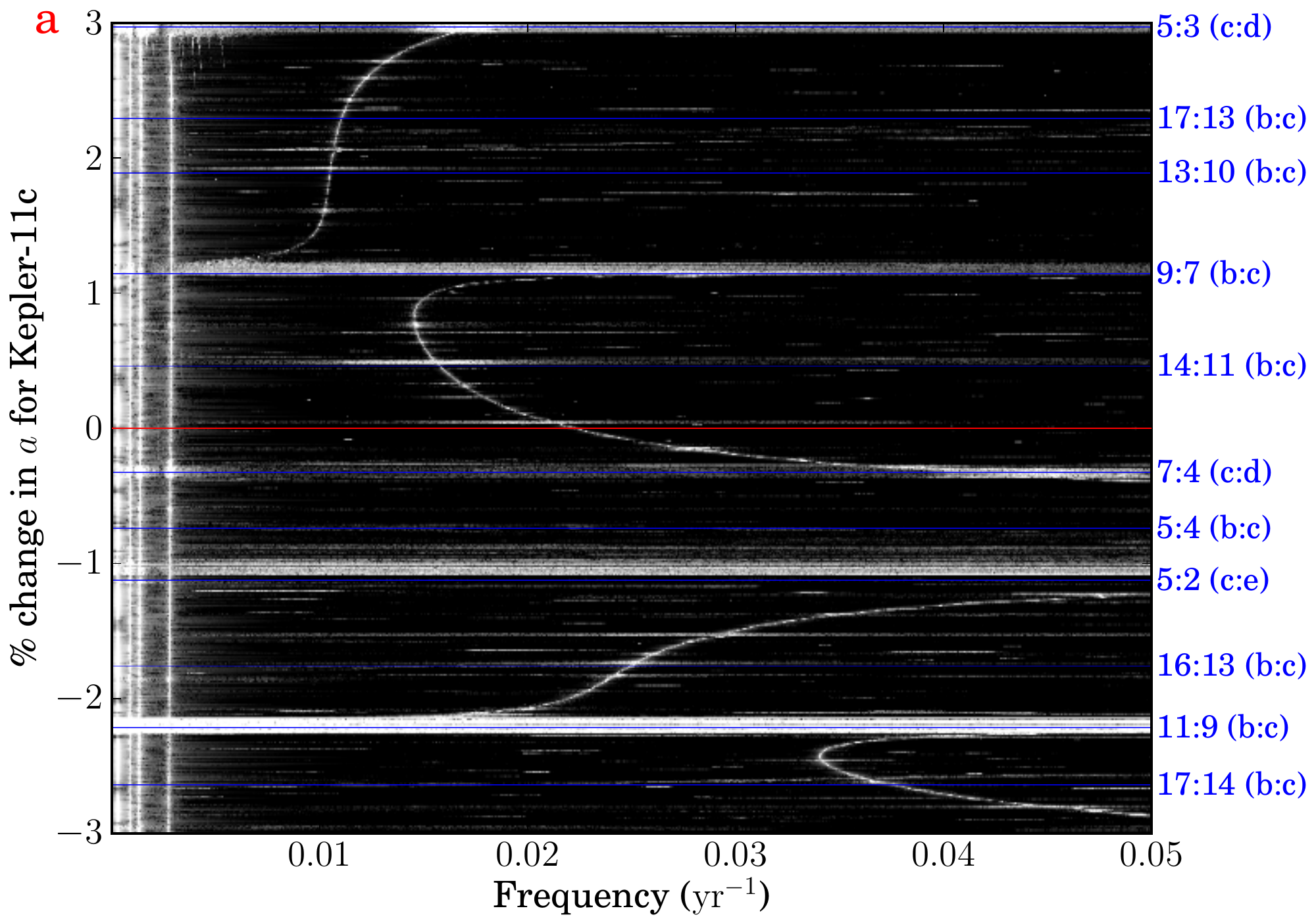}
	\includegraphics[width=0.495\textwidth,trim=0 0 0 0,clip]{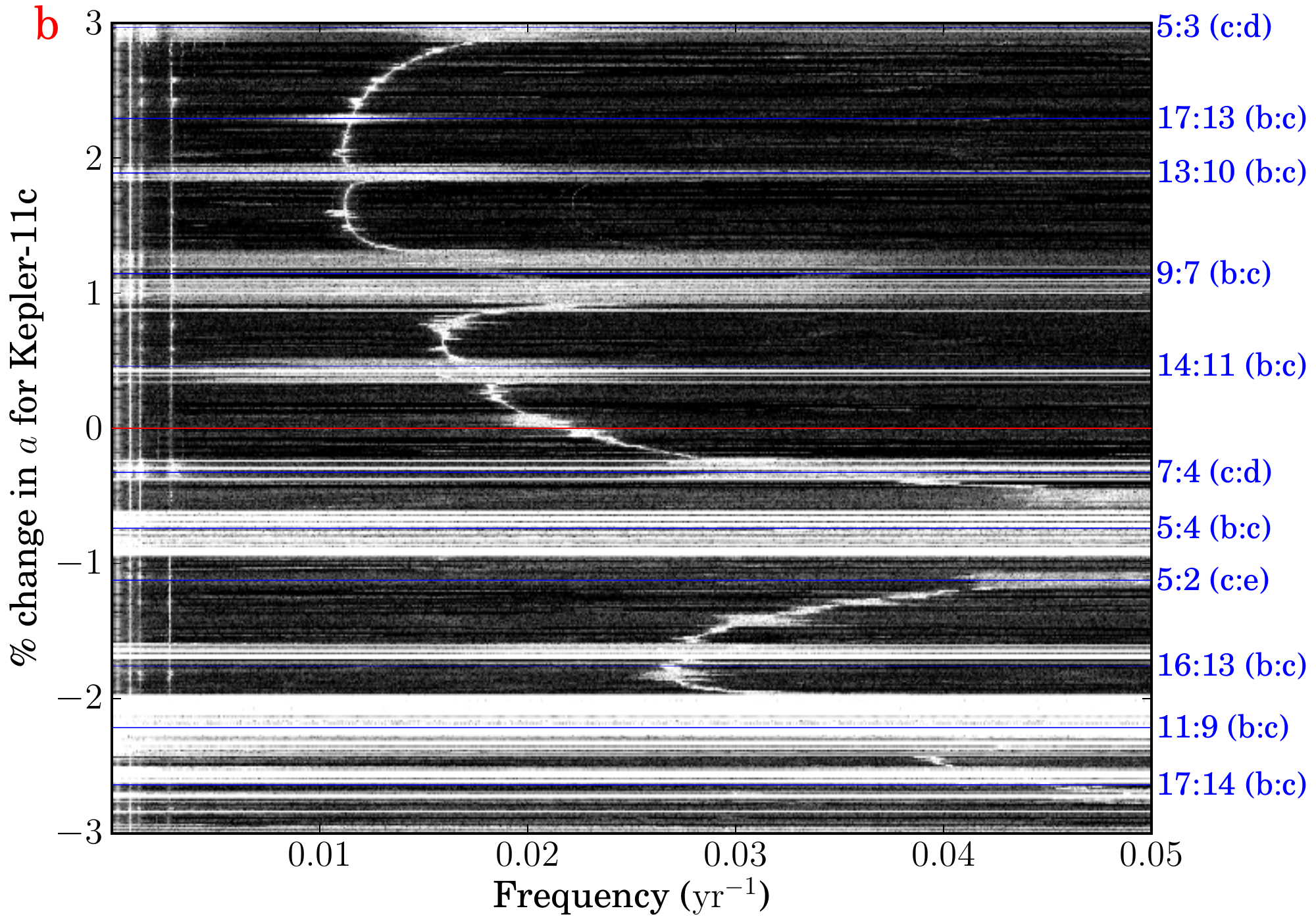}\\
	\includegraphics[width=0.495\textwidth,trim=0 0 0 0,clip]{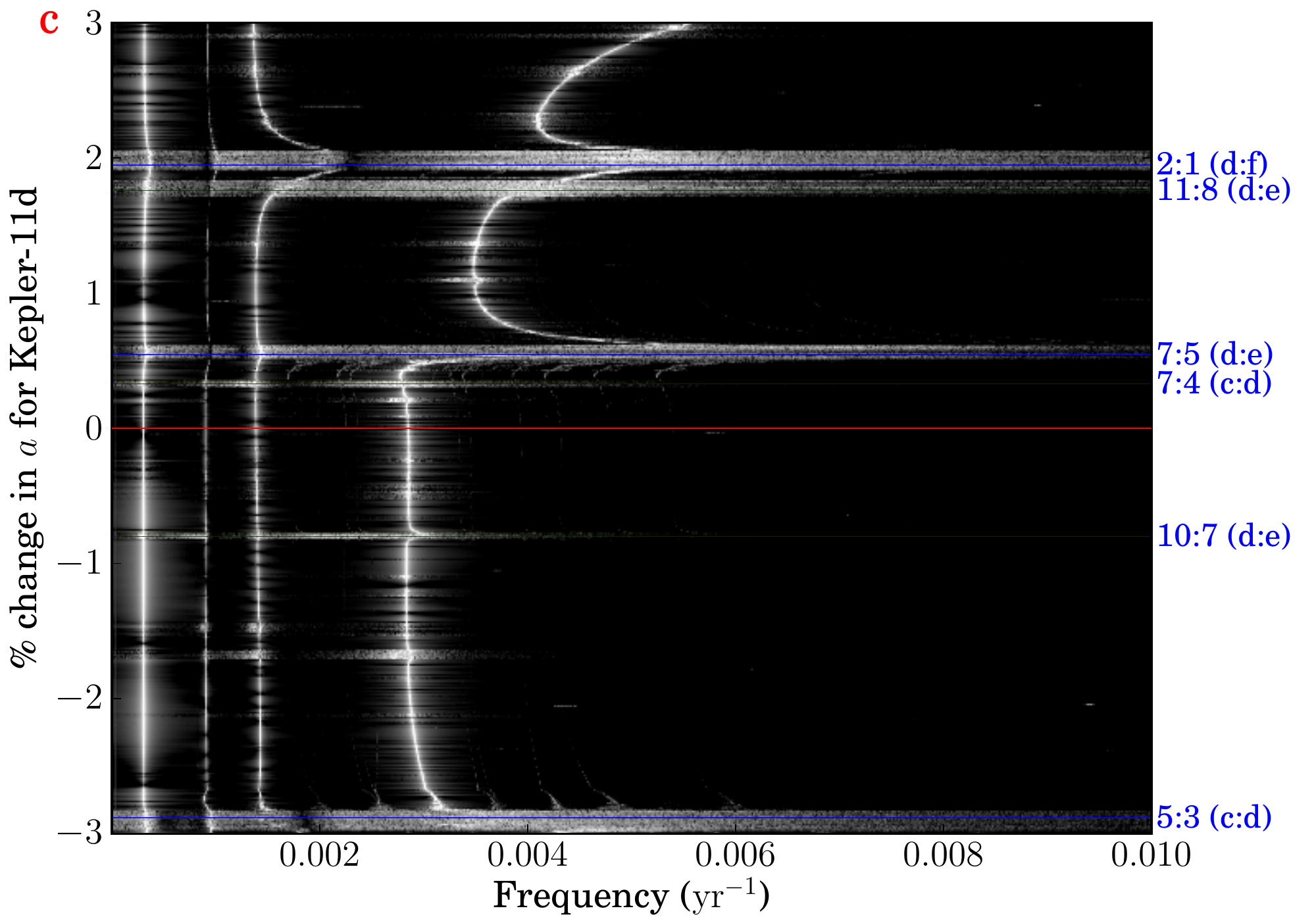}
	\includegraphics[width=0.495\textwidth,trim=0 0 0 0,clip]{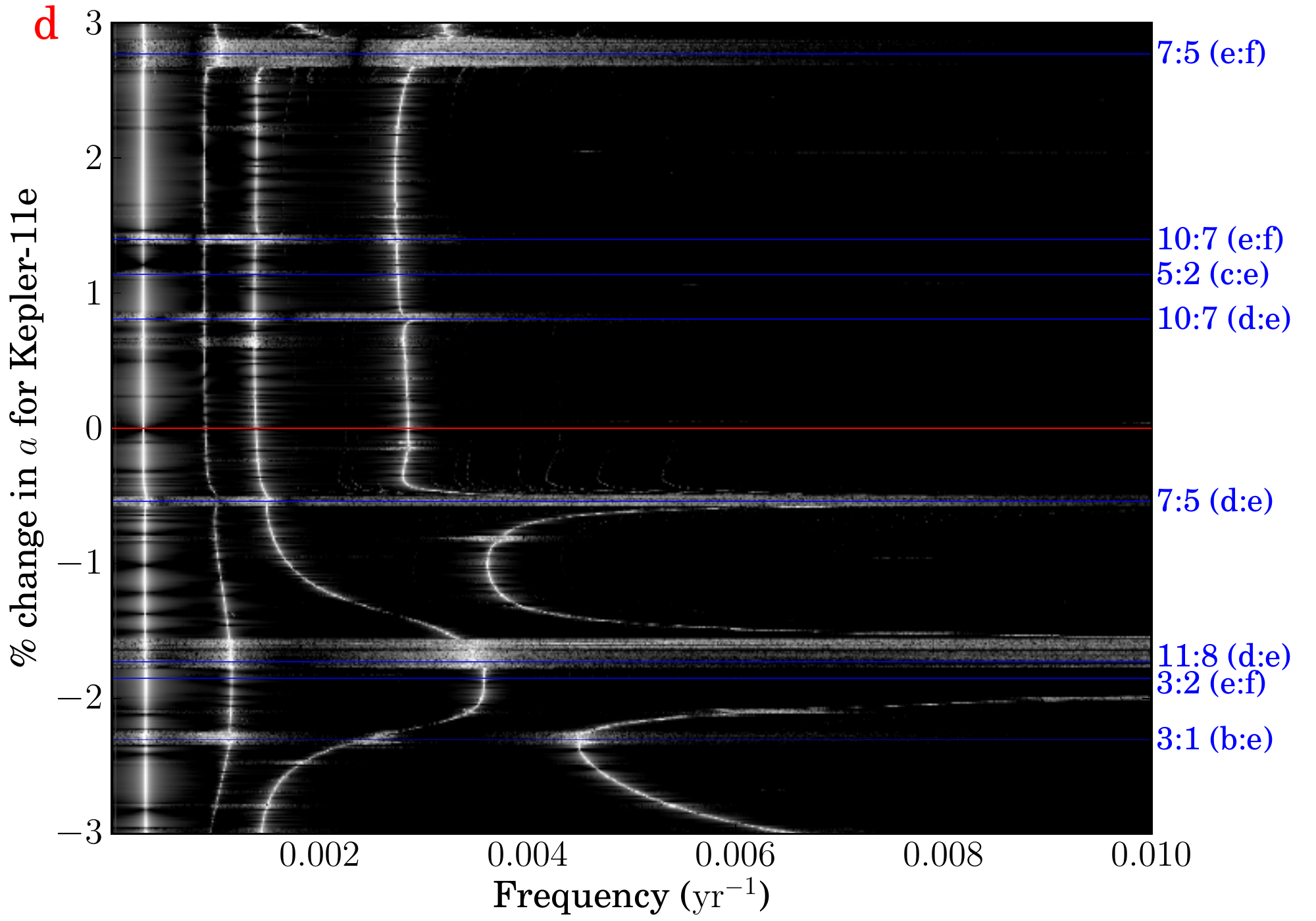}\
	\caption{\footnotesize Frequency maps of the Kepler-11 planets (c, d and e here) when the semi-major axis for the planet in question is altered by up to $3\%$ around its standard value (indicated by the red lines). In panel b, we also boost $e_c$ to $0.028$, $10x$ the value as in Table \ref{table:elements}, to demonstrate the dependence of instability on eccentricity. Each blue line designates a 2-body mean-motion resonances ($p:q$) between a pair of planet -- we label all resonances with $|p-q| \leq 3$. Additionally, the fourth-order 17:13 (b:c) resonance is labeled in plots (a) and (b) to show how a higher eccentricity causes higher-order resonances to become more chaotic. The colormap is logarithmic spanning two orders of magnitude in Fourier amplitude from black to white. Strongly chaotic/unstable locales (show up as white horizontal bands) are typically associated with a low-order MMR, though the reverse is not necessarily true. Instability in panel b is much more wide-spread, as is expected given its higher eccentricity. For our definition of dynamical stability, the fraction of chaotic/unstable systems in the four panels are, respectively, (a) $13 \%$, (b) $53 \%$, (c) $17 \%$, and (d) $16 \%$.}
	\label{fig:sm-a}
\end{figure*}

In this set of experiment, the semi-major axis of one of the six planets is varied by up to $3\%$ from its standard value. Each panel in Fig. \ref{fig:sm-a} is a compilation of $601$ simulations each integrated for a period of $5.24$ Myrs, corresponding to $\sim 2\times 10^8$ orbital periods of Kepler-11b. The number of simulations gives us a resolution in semi-major axis space of $da/a \sim 10^{-4}$. Only cases where we vary the axis of $c,d$ and $e$ are shown. In addition, we present a case where we boost the eccentricity of $c$ to $0.028$, $10$ times its listed value.

As Fig. ~\ref{fig:sm-a} shows, the observed Kepler-11 system (indicated by red lines) is stable, a fact also confirmed by our very long integration (250 Myrs). However, in its immediate neighbourhood, many unstable bands exist.  For instance, moving Kepler-11c outward by a mere $0.05\%$ will bring the system into an unstable region. Regions that show up as wide and bright horizontal stripes have (or will soon have) close encounters within our short integration. 

Using these visual images of instabilities, we proceed to discuss the following topic: is the system finely tuned for stability? and what causes the instability?

Given our working definition, the fractions of unstable regions in Fig. \ref{fig:sm-a} are $13-17\%$ at low eccentricity (Table \ref{table:elements}), jumping sharply to $53\%$ when $e_c$ is boosted to $0.028$.\footnote{This latter high fraction explains why many of the systems investigated by \citet{Migas,LissauerNature} are unstable: the system can only tolerate very low eccentricity values.}  This argues that, if the real system has eccentricities comparable to those in the standard case, it is not overly fine-tuned. Much of the phase space around the standard case is stable. This runs against the conclusion in \citet{Migas} who argued that their stable orbital solution (case IVa) is very fragile and is enveloped by a dense web of chaotic regions, brought about by a swarm of 3-body resonances. This difference in conclusion may arise from our adoption of a different fast indicator than that in \citet{Migas}. It is possible that nearly $100\%$ of the phase space is indeed criss-crossed by 3-body resonances and may be formally chaotic (with a positive Lyapunov exponent), but it has little long-term dynamical effects. The system, when analyzed in frequency space, behaves largely quasi-periodically, at least for most of the phase space around the standard case. This is akin to the case of the outer solar system, which, although criss-crossed by many 3-body resonances \citep{Guzzo05,Guzzo06} and exhibits short Lyapunov exponents, enjoys exceedingly long dynamical lifetime, perhaps up to $10^{18}$ yrs \citep{MurrayHolman}.

Now we turn to the origin of instability. In panel a of Fig. \ref{fig:sm-a}, we delineate in blue lines the locations of all low-order 2-body MMRs that Kepler-11c can be engaged in (up to order $3$). The neighbourhood is peppered by these resonances. This property is also shared by other planets of the system, except for Kepler-11g as it lies much further from the rest to be near low-order resonances. The proximity to these resonances is expected for a closely-packed high-multiplicity planetary system like Kepler-11. What is surprising, however, is the fact that unstable regions are more often than not associated with these 2-body MMRs. It is also interesting to observe that some secular frequencies undergo dramatic detours near some particularly destabilizing MMRs. This detour effect is understood \citep{BrouwerClemence,Laskar88}, but its correlation with unstable regions is not known. In \S \ref{subsec:speculate}, we speculate on how this may explain instabilities in multiple planetary systems.

The effects of 2-body MMRs are not straightforward to understand. While regions around some first-order 2-body resonances are not destabilized (such as the 5:4 (b:c) and 3:2 (e:f) MMRs), or that the unstable bands are too narrow in $a$-space to be resolved by our simulations, others are (the 2:1 (d:f) resonance). Second- and third-order resonances appear to be just as effective as first-order MMRs in causing havoc. Even higher order 2-body MMRs may account for some of the unstable regions that are not associated with low-order MMRs. \citet{Migas,Quillen} have argued that, given the ubiquity of 3-body MMRs, they are dynamically important for the stability of multiple-planet systems. However, most of our unstable regions are associated with 2-body MMRs. We argue that either 3-body MMRs are too weak to be dynamically important, or they act through the hands of 2-body MMRs.

We now comment on the effect of a higher eccentricity. The unstable fraction in the frequency map rises steeply with planet eccentricity. In panel b of Fig. ~\ref{fig:sm-a}, when the eccentricity of Kepler-11c is increased ten-fold (to $0.028$), each unstable zone widens substantially and new unstable regions appear (or, widens to become resolvable by our simulations). This is caused by the growth in resonance strength as eccentricity increases. Thus low eccentricities are essential for the stability of closely-packed systems while higher eccentricities, even at a few percent level, strongly affect system stability. In the following, we further quantify this dependence.

\subsection{Eccentricity space}

\begin{figure*}
    \centering
    \includegraphics[width=0.495\textwidth,trim=0 0 0 0,clip]{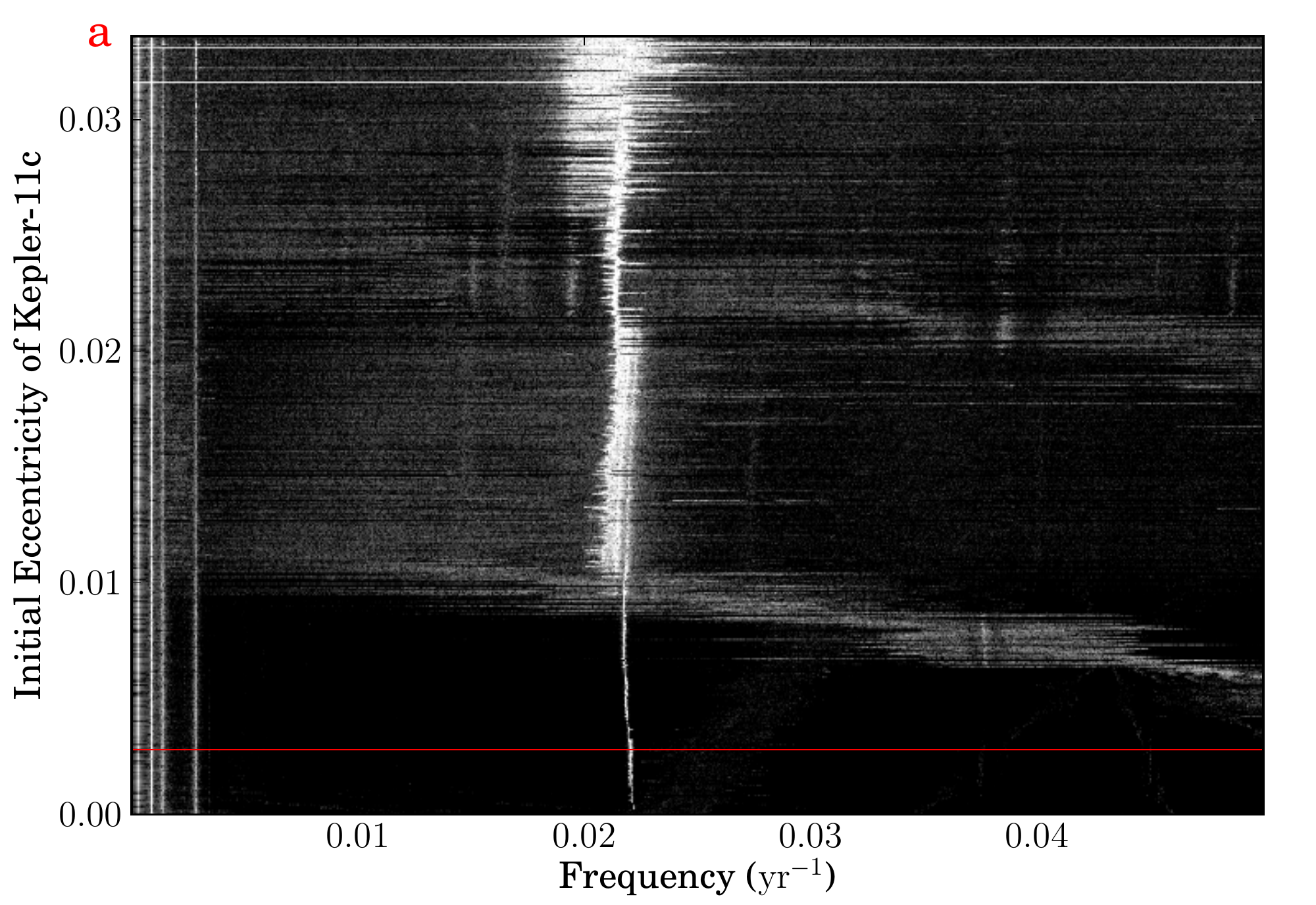}
    \includegraphics[width=0.495\textwidth,trim=0 0 0 0,clip]{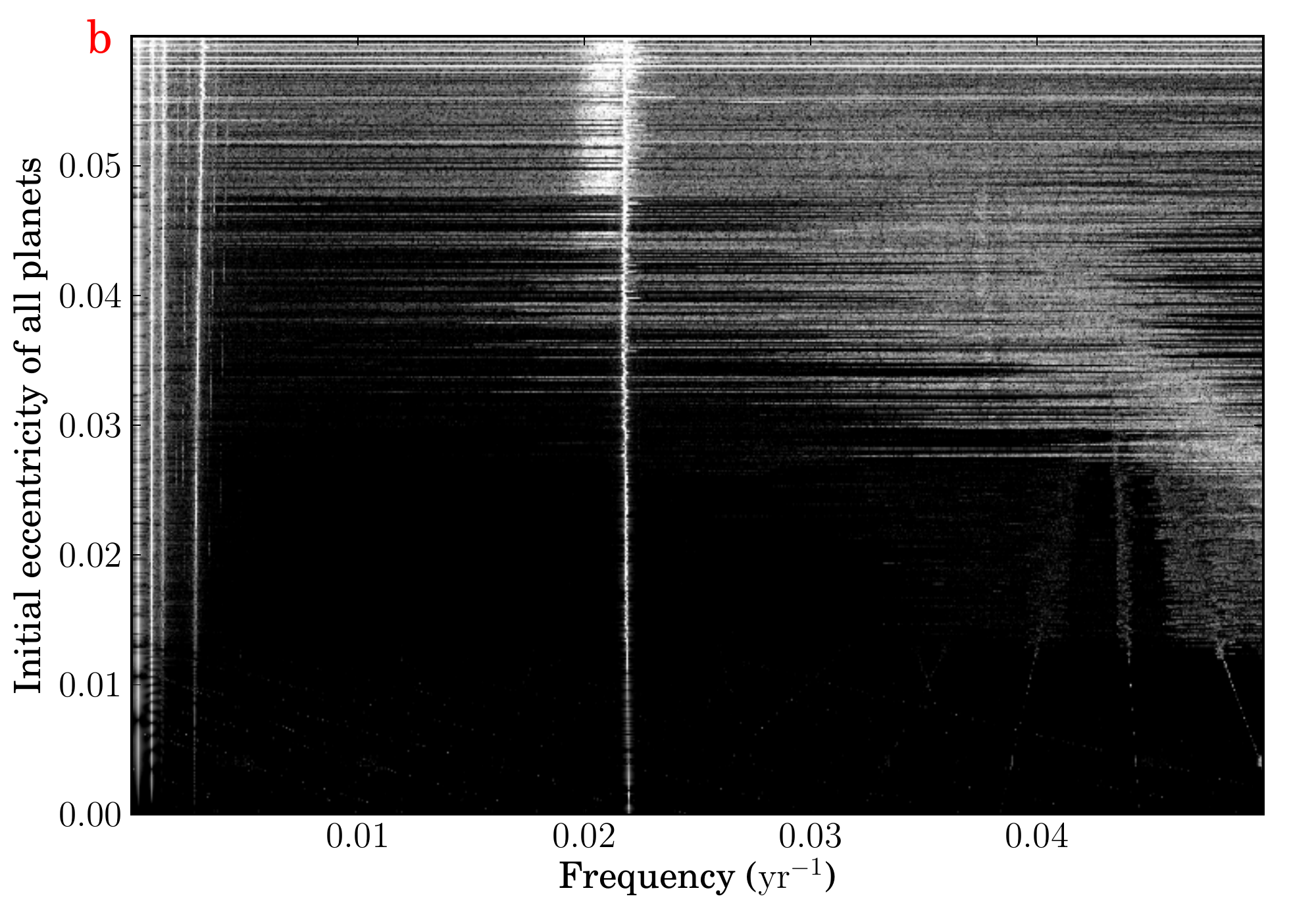}\
    \caption{\footnotesize  Frequency maps, obtained by varying either the eccentricity of Kepler-11c alone (panel a) or those of all planets (panel b, same eccentricity for all planets). The former frequency map is generated from $z_c$, while the latter one from $z_b$. As eccentricities rise, instabilities become more prevalent. Hardly any stable systems remain beyond eccentricity of a few percent.}
    \label{fig:sm-e}
\end{figure*}

At the observed periods (and semi-major axis), the Kepler-11 system avoid low-order MMRs and is stable at low eccentricities. Here we explore how this stability is destroyed at high eccentricities. Each panel of Fig. \ref{fig:sm-e} is again a compilation of 601 simulations, run for $5.24$ Myr each, with panel a showing the impact of raising $e_c$ alone, and panel b that of raising eccentricities of all planets (assumed same for all planets). 

It appears that rising eccentricity alters the dynamics gradually: secular frequencies are mildly modified; their amplitudes become higher; their harmonics begin to appear; power spill-over to neighbouring frequencies become increasingly prominent. These are all manifestations of nonlinear secular interactions. One observes that instabilities tend to occur when one of the secular frequencies (or its low-order harmonics) intersect with other secular frequencies. This suggests that the instabilities we witness may be related to secular chaos, chaos produced by overlapping secular resonances. In the case of the Solar system, even with values of eccentricity/inclination at a few percent, secular chaos in the inner region will lead eventually to the ejection of Mercury \citep{Laskar90,Lithwick2011}. For the Kepler-11 system, Fig. \ref{fig:sm-e} suggests that full scale chaos also occurs at a similar threshold eccentricity: for our working definition, this is $e\sim 0.04$.

Panel a of Fig. \ref{fig:sm-e} documents the change in frequency map as $e_c$ alone is raised.  A stable region exists for $e_c < 0.01$. A weakly chaotic region, in which systems become progressively less periodic but still manage to maintain long-term stability, can be seen for $0.01 < e_c < 0.03$. In this region, small shifts in planets' position from their current values will very likely lead to dynamical havocs, as is shown in panel b of Fig. \ref{fig:sm-a}. Above $e_c \sim 0.03$, instability occurs even at the observed positions.

Similar analysis by raising the eccentricities of the other planets yields similar upper bound for long-term stability, with the exception of Kepler-11f. This planet, being relatively light, has an upper bound of $e_f \sim 0.09$. If all the planets are given the same eccentricity instead (Figure~\ref{fig:sm-e}b), the stability threshold lies at $e \sim 0.04$.

If, according to \citet{Lissauer2013}, the inner pair of planets do have lower masses ($M_b < 2.2 M_\Earth$, $M_c < 4 M_\Earth$), then the threshold eccentricity is higher. We have verified that the new \citet{Lissauer2013} solution (low mass, high eccentricity) is stable for 200 Myrs.

We conclude that the Kepler-11 system must have near-circular orbits (eccentricities smaller than a few percent) to be long-term stable. This constraint may restrict the formation scenario of the system, as we discuss below.

\subsection{Planet Masses}

\begin{figure*}
    \centering
    \includegraphics[width=0.495\textwidth,trim=0 0 0 0,clip]{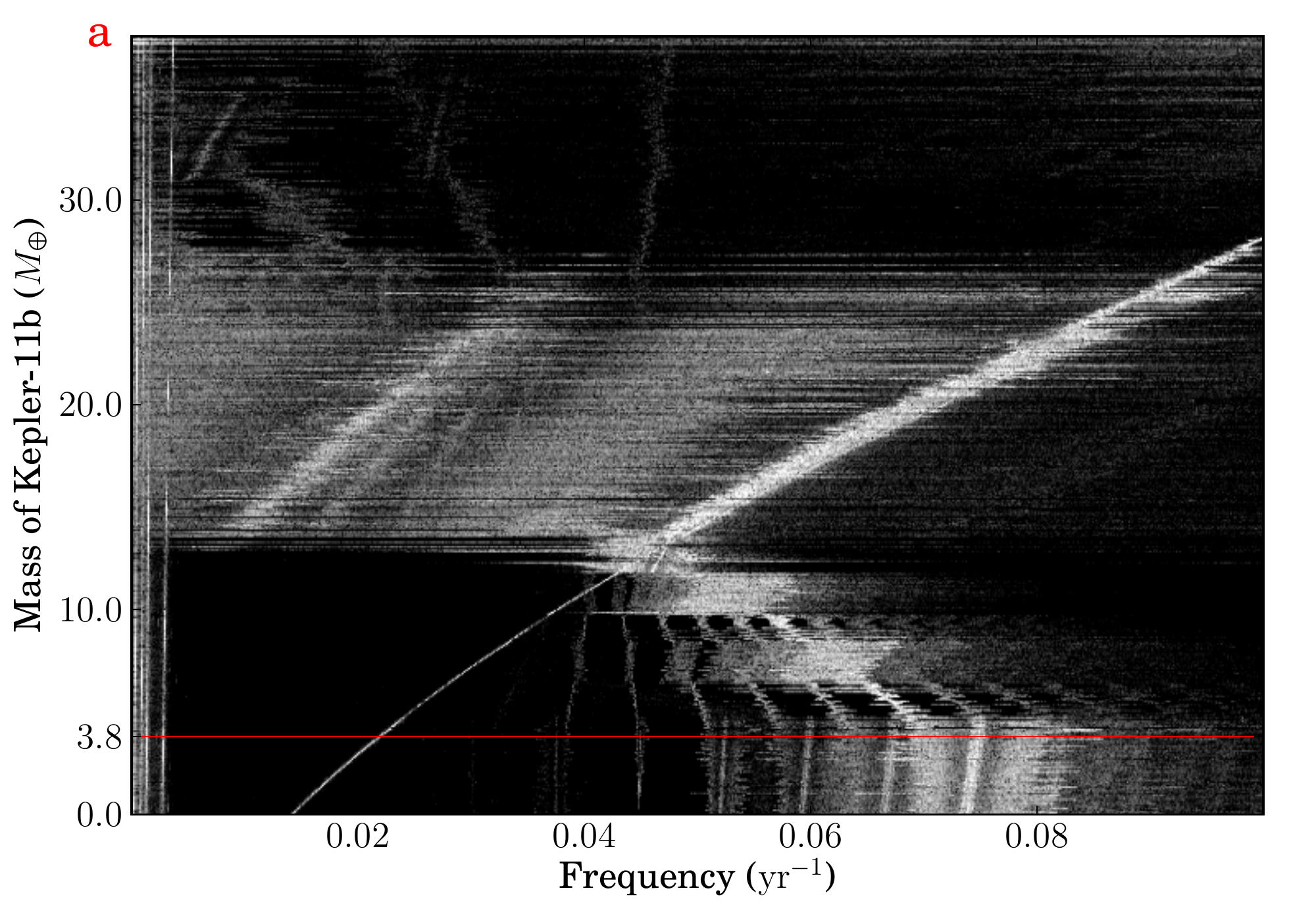}
    \includegraphics[width=0.495\textwidth,trim=0 0 0 0,clip]{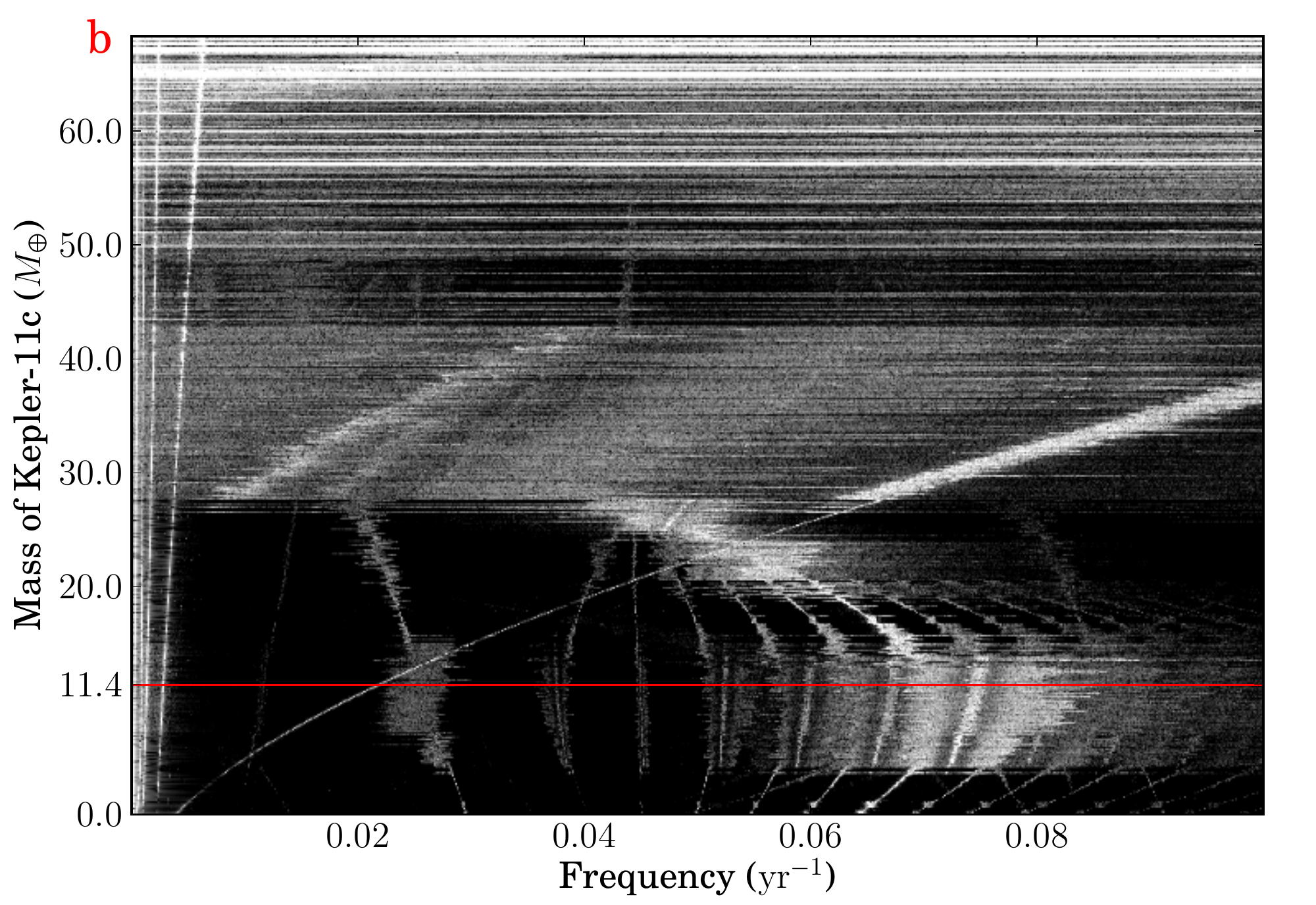}\
    \caption{\footnotesize Frequency map when mass of Kepler-11b is varied (panel a), and when mass of Kepler-11c is varied (panel b). The red lines designate their current masses (Table~\ref{table:elements}). Secular frequencies (and their harmonics) are affected by the varying planet masses. But the system remains stable as long as the masses remain below a few times their standard values.}
    \label{fig:sm-m}
\end{figure*}

We explore the effects of changing planet masses on stability, again by varying the mass of a specific planet and integrating for $5.24$ Myrs. We focus on the inner pair as their envelopes are the most likely to have been evaporated by stellar irradiation \citep{Lopez,OwenWu}. We find that, if the eccentricities are as low as those in Table \ref{table:elements}, masses for planets b and c can be three to four times higher than their fiducial values, while the upper bounds for masses of the outer planets are all above $20 M_\Earth$. All current mass determinations \citep{Lissauer2013,Migas} yield values below these limits. So we can not exclude the possibility that significant mass-loss (caused by, e.g., photoevaporation) has occurred in the past.

\subsection{Speculation on the Origin of instability}
\label{subsec:speculate}

What is the origin of orbital instability in multiple planet systems like Kepler-11?  Despite a number of numerical studies \citep{Chambers,Duncan,Smith,Funk}, this origin remains obscure. Possibilities include secular interaction, overlapping 2-body MMRs, or overlapping 3-body MMRs \citep{Quillen}. We have not attempted to answer this question systematically, but discuss an interesting possibility suggested by our analysis of the frequency maps.

Near regions of instability (typically coincide with MMRs, Fig. \ref{fig:sm-a}), we often observe that the secular frequencies are strongly varying. This has the prospect of sweeping through secular resonances -- commensurability in frequencies between different secular modes. We speculate that a combination of MMRs and secular interactions is responsible for instability in systems resembling Kepler-11 and plan to investigate this further in the future.

\section{Origin and history of the system}

We now discuss how stability may cast constraints on the formation history of Kepler-11.

In the {\it in situ} formation scenario, championed by \citet{HansenMurray,ChiangLaughlin}, the Kepler-11 system is assumed to be assembled locally by agglomerating planetesimals, similar to how the terrestrial planets are thought to have formed \citep{ChambersWetherill,Chambers01}. In these latter studies, terrestrial planets typically acquire eccentricities of $0.1-0.2$, due to close-encounters and collisions among the proto-planets \citep[e.g.][]{Raymond06}. This value is roughly the ratio of the surface escape velocity to the local Keplerian velocity. Terrestrial eccentricities may then be damped down by residual planetesimals to their modern values. We believe such a course could not have taken place in the Kepler-11 system -- Fig. \ref{fig:sm-a}b shows that, even with a $e_c$ as low as $0.028$, $53\%$ of the neighbourhood around planet $c$ is unstable; when planet eccentricities reach upward of $0.04$, hardly any stable systems exist in the observed neighbourhood. This argues that the Kepler-11 system, if formed by conglomeration, has to stay low in eccentricity. This requires excess dissipation over that in the scenario of terrestrial formation.

Many works have proposed that {\it Kepler} planets are migrated to their current positions \citep[see,e.g.][]{TerquemPap,pap2005,ogihara13}. Our stability analysis challenges this proposal. Fig.~\ref{fig:sm-a} shows that the current stable state of Kepler-11 is sandwiched between chaotic regions -- widening or compressing the planet spacing away from their current values, one is guaranteed to encounter a chaotic band. So either the system is migrated inward {\it en masse}, preserving the current period ratios, or the system is assembled through differential migration that somehow manages to cross the many MMRs safely -- `cross' and `safely' are the two key words here. We investigate whether this is possible.

\subsection{Migration past MMRs}
\label{subsec:migration}

If planets are migrated convergingly, they will encounter multiple MMRs which raise their eccentricities and lead to orbit crossing. For stability, migration has to be accompanied by damping of eccentricity. We define the migration and $e$-damping timescale as
\begin{equation}
\label{eq:eq1}
\tau_a \equiv - \frac{a}{\dot{a}}\,, \qquad \tau_e \equiv - \frac{e}{\dot{e}}\,,
\end{equation}
where in our simulation, we simply link the two by a constant, $\tau_e = \tau_a/K$. To implement these into the SWIFTER code, we adopt the following form of acceleration for planet migration \citep[c.f.][]{GoldreichSchlichting}
\begin{equation}
\vec{a} = \alpha\vec{r} + \beta\vec{v}\, ,
\end{equation}
where $\vec{v}$ and $\vec{r}$ are the velocity and position vectors of the planet. Using the fact that angular momentum variation $d\vec{L}/dt = \vec{r}\times {\vec{a}}$, and energy variation $dE/dt = \vec{v} \cdot{\vec{a}}$, we obtain the expressions for the acceleration:
\begin{align}
\alpha = \frac{-K e^2 ({\vec{v}\cdot\vec{v}})}{(1-e^2) (\vec{r}\cdot\vec{v}) {\tau}_{a}}\,, \qquad
\beta = \frac{K e^2}{(1-e^2) {\tau}_{a}} - \frac{1}{2 {\tau}_{a}}\, ,
\end{align}

\begin{figure}
	\centerline{\includegraphics[width=0.49\textwidth,trim=0 0 0 0,clip]{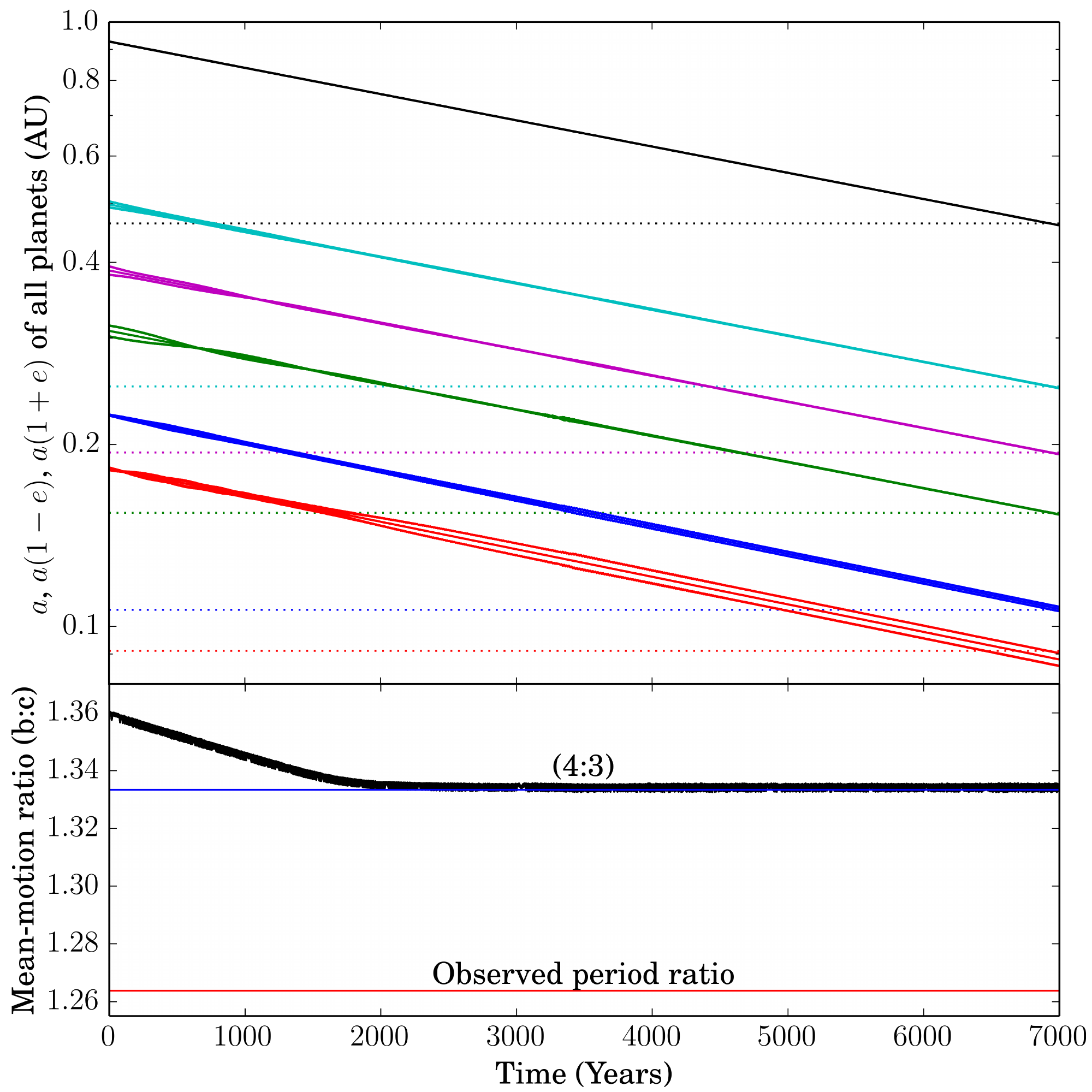}}
	\caption{\footnotesize Planet migration simulation using $K=10$ and $\tau_a = 10^4$ yrs. Planets are initialized at twice their current semi-major axes, with Kepler-11c getting an additional $5\%$ and a smaller $\tau_a$. This allows it to cross some MMRs as it approaches its current position. The top panel shows the migration of each planet, while the bottom panel shows the mean-motion ratio between planets b and c. The pair gets trapped in the 4:3 resonance during migration. The dashed lines in the top panel represent the current semi-major axes of the planets.}
	\label{fig:trap}
\end{figure}

With this simple formulation, we migrate all planets from twice their current semi-major axis to their current values. We allow for only a small amount of differential migration by displacing the position of a particular planet by $5\%$ inward (or outward) relative to the other planets. Its $\tau_a$ is adjusted correspondingly so that it can reach its current position in sync with its siblings. As it does so, the planet would need to cross some MMRs with its neighbours and we observe how the MMRs, destabilizing in \S \ref{sec:stability}, affects the migration.

Fig. \ref{fig:trap} shows an example simulation with $\tau_a = 10^4$ yrs, the expected value for Type I migration in this neighbourhood. Planet c is moved outward by $5\%$ relative to the other ones and the simulation is truncated at $8000$ yrs. We adopt $K=10$, though results for $K=100-1000$ look similar. Since much of the instability exhibited in Fig. \ref{fig:sm-a} occurs in $10^5 - 10^6$ yrs timescale, the rapid dissipation we introduce here ($\tau_e = 10^3$ yrs) stabilizes the system and it is able to migrate safely across a number of weak MMRs. However, relative migration of the inner pair is stalled as they encounter the 4:3 MMR. This also occurs when we artificially displace other planets -- the 4:3 MMR between planets b and c, the 3:2 MMR between d an e, and the 3:2 MMR between e and f are the three strongly trapping MMRs in this system.

Differential migration leading to resonance trapping is observed by multiple other studies \citep[see, e.g.][]{TerquemPap}. This contrasts strongly against the observed paucity of MMR pairs among {\it Kepler} planets, and casts doubt on the plausibility of the migration proposal. \citet{GoldreichSchlichting} tried to dispell this doubt by demonstrating, analytically, that resonance trapping will not happen for sufficiently weak eccentricity damping (small $K$).  For the 4:3 MMR, their analytical criterion for resonance escape is $K \leq 100 (\mu/3.6\times 10^{-5})^{-2/3}$, where we have scaled the planet mass by that of Kepler-11c. Why do we observe resonance trapping, even with $K$ as small as $10$?  The reason lies in our particular set-up for the migration. The migration rate relevant for resonance trapping is the rate of differential migration, which in our case turns out to be $20$ times smaller than the individual migration rate. As a result, the effective $K$ in our simulation is $K\times 20 = 200$. This explains why we observe resonance trapping. Adopting higher values of $K$ will lead to even more secure trapping. So our simulation of small differential migration always leads to resonance trapping.

For type I migration, one expects $\tau_a$ to scale linearly with planet mass. This indicates strong differential migration for the planets in Kepler-11.  What will happen were we to adopt such a large differential migration?  In this case, resonance trapping may be avoided \citep{GoldreichSchlichting}, but unfortunately, there is also no stopping for the planets to crash into each other. 

So we are left with little option to accommodate the migration scenario: if differential migration is very small, the planets will be trapped in resonances; if it is large, the planets will encounter each other within a short time. The Kepler-11 system could have migrated while maintaining their current period ratios, but that seems an implausible coincidence. Stochastic migration \citep{rein12}, in which planets are randomly disturbed, is also hard to accommodate -- the planets are currently too well packed to allow much random shuffling; and the upper limit on planet eccentricity is too strict to allow much meddling.

\section{Summary}

The Kepler-11 system is a high-multiplicity, closely-packed system. We study the dynamics of the system through `frequency maps', the map of Fourier transform for the complex eccentricity vectors. We use frequency maps generated by short integrations (typically $5$ Myrs) to predict the long term stability of the system, in contrast to previous works which use Lyapunov exponents.

We find that the current system \citep[with parameters as determined by][]{LissauerNature} is stable. In semi-major axis space, the current system is sandwiched between unstable bands, the latter having a measure of $13-17\%$. So the stability is not as fragile as claimed by previous studies. In eccentricity space, we find the stability is erupted when planet eccentricities rise above $0.04$. In mass space, we find that the masses of planets Kepler-11b and 11c can be a few times higher while still maintaining system stability. This allows for the possibility that these planets have suffered substantial mass-loss in the past, due to their proximity to the host star.

The frequency maps also lead us to speculate on the origin of dynamical instability in closely packed systems. We find that chaotic/unstable bands in semi-major axis are typically associated with low-order mean-motion resonances between 2 planets. Three-body resonances, proposed in the literature to explain the instability, are not conspicuously evident in the frequency maps. If they are indeed dynamically important, they would have to act through the hands of 2-body resonances. We suggest, instead, that secular dynamics coupled with mean-motion resonances may be the real culprit. More study in the future is needed.

The stability study also provides constraints on the assembly history of Kepler-11. {\it In situ} formation is possible, if planet eccentricities are maintained low at all times. Assembly by migration is hard to accommodate: even with strong eccentricity damping, migration has to finely tuned to avoid either trapping into mean motion resonances, or planets crashing into each other.

The Kepler-11 system is likely not unique. The {\it Kepler} mission has disclosed that a substantial fraction of planetary systems are similarly closely-packed, nearly coplanar and circular.  Further refinement on system parameters, and more extensive exploration for the formation path of this kind of system will prove fruitful for understanding planet formation.

\acknowledgements NM acknowledges NSERC for an undergraduate summer research award. YW's research is supported by NSERC and the province of Ontario.


\begin{thebibliography}{46}
\expandafter\ifx\csname natexlab\endcsname\relax\def\natexlab#1{#1}\fi

\bibitem[{{Agol} {et~al.}(2005){Agol}, {Steffen}, {Sari}, \& {Clarkson}}]{Agol}
{Agol}, E., {Steffen}, J., {Sari}, R., \& {Clarkson}, W. 2005, \mnras, 359, 567

\bibitem[{{Brouwer} \& {Clemence}(1961)}]{BrouwerClemence}
{Brouwer}, D. \& {Clemence}, G.~M. 1961, {Methods of celestial mechanics}

\bibitem[{{Chambers}(2001)}]{Chambers01}
{Chambers}, J.~E. 2001, \icarus, 152, 205

\bibitem[{{Chambers} \& {Wetherill}(1998)}]{ChambersWetherill}
{Chambers}, J.~E. \& {Wetherill}, G.~W. 1998, Icarus, 136, 304

\bibitem[{{Chambers} {et~al.}(1996){Chambers}, {Wetherill}, \&
  {Boss}}]{Chambers}
{Chambers}, J.~E., {Wetherill}, G.~W., \& {Boss}, A.~P. 1996, \icarus, 119, 261

\bibitem[{{Chiang} \& {Laughlin}(2013)}]{ChiangLaughlin}
{Chiang}, E. \& {Laughlin}, G. 2013, \mnras, 431, 3444

\bibitem[{{Duncan} \& {Lissauer}(1997)}]{Duncan}
{Duncan}, M.~J. \& {Lissauer}, J.~J. 1997, \icarus, 125, 1

\bibitem[{{Franklin} {et~al.}(1993){Franklin}, {Lecar}, \&
  {Murison}}]{Franklin}
{Franklin}, F., {Lecar}, M., \& {Murison}, M. 1993, \aj, 105, 2336

\bibitem[{{Froeschle}(1984)}]{Froes}
{Froeschle}, C. 1984, Celestial Mechanics, 34, 95

\bibitem[{{Froeschl{\'e}} {et~al.}(1997){Froeschl{\'e}}, {Lega}, \&
  {Gonczi}}]{Froes97}
{Froeschl{\'e}}, C., {Lega}, E., \& {Gonczi}, R. 1997, Celestial Mechanics and
  Dynamical Astronomy, 67, 41

\bibitem[{{Funk} {et~al.}(2010){Funk}, {Wuchterl}, {Schwarz}, {Pilat-Lohinger},
  \& {Eggl}}]{Funk}
{Funk}, B., {Wuchterl}, G., {Schwarz}, R., {Pilat-Lohinger}, E., \& {Eggl}, S.
  2010, \aap, 516, A82

\bibitem[{{Goldreich} \& {Schlichting}(2014)}]{GoldreichSchlichting}
{Goldreich}, P. \& {Schlichting}, H.~E. 2014, \aj, 147, 32

\bibitem[{{Go{\'z}dziewski} {et~al.}(2001){Go{\'z}dziewski}, {Bois},
  {Maciejewski}, \& {Kiseleva-Eggleton}}]{Goz}
{Go{\'z}dziewski}, K., {Bois}, E., {Maciejewski}, A.~J., \&
  {Kiseleva-Eggleton}, L. 2001, \aap, 378, 569

\bibitem[{{Guzzo}(2005)}]{Guzzo05}
{Guzzo}, M. 2005, \icarus, 174, 273

\bibitem[{{Guzzo}(2006)}]{Guzzo06}
---. 2006, \icarus, 181, 475

\bibitem[{{Guzzo} {et~al.}(2002){Guzzo}, {Kne{\v z}evi{\'c}}, \&
  {Milani}}]{Guzzo}
{Guzzo}, M., {Kne{\v z}evi{\'c}}, Z., \& {Milani}, A. 2002, Celestial Mechanics
  and Dynamical Astronomy, 83, 121

\bibitem[{{Hansen} \& {Murray}(2011)}]{HansenMurray}
{Hansen}, B.~M.~S. \& {Murray}, N. 2011, ArXiv e-prints

\bibitem[{{Holman} \& {Murray}(2005)}]{HolmanMurray}
{Holman}, M.~J. \& {Murray}, N.~W. 2005, Science, 307, 1288

\bibitem[{{Ito} \& {Tanikawa}(2002)}]{Ito}
{Ito}, T. \& {Tanikawa}, K. 2002, \mnras, 336, 483

\bibitem[{{Laskar}(1988)}]{Laskar88}
{Laskar}, J. 1988, \aap, 198, 341

\bibitem[{{Laskar}(1990)}]{Laskar90}
---. 1990, \icarus, 88, 266

\bibitem[{{Laskar} {et~al.}(1992){Laskar}, {Froeschl{\'e}}, \&
  {Celletti}}]{Laskar92}
{Laskar}, J., {Froeschl{\'e}}, C., \& {Celletti}, A. 1992, Physica D Nonlinear
  Phenomena, 56, 253

\bibitem[{{Lissauer} {et~al.}(2011){Lissauer}, {Fabrycky}, {Ford}, {Borucki},
  {Fressin}, {Marcy}, {Orosz}, {Rowe}, {Torres}, {Welsh}, {Batalha}, {Bryson},
  {Buchhave}, {Caldwell}, {Carter}, {Charbonneau}, {Christiansen}, {Cochran},
  {Desert}, {Dunham}, {Fanelli}, {Fortney}, {Gautier}, {Geary}, {Gilliland},
  {Haas}, {Hall}, {Holman}, {Koch}, {Latham}, {Lopez}, {McCauliff}, {Miller},
  {Morehead}, {Quintana}, {Ragozzine}, {Sasselov}, {Short}, \&
  {Steffen}}]{LissauerNature}
{Lissauer}, J.~J., {Fabrycky}, D.~C., {Ford}, E.~B., {Borucki}, W.~J.,
  {Fressin}, F., {Marcy}, G.~W., {Orosz}, J.~A., {Rowe}, J.~F., {Torres}, G.,
  {Welsh}, W.~F., {Batalha}, N.~M., {Bryson}, S.~T., {Buchhave}, L.~A.,
  {Caldwell}, D.~A., {Carter}, J.~A., {Charbonneau}, D., {Christiansen}, J.~L.,
  {Cochran}, W.~D., {Desert}, J.-M., {Dunham}, E.~W., {Fanelli}, M.~N.,
  {Fortney}, J.~J., {Gautier}, III, T.~N., {Geary}, J.~C., {Gilliland}, R.~L.,
  {Haas}, M.~R., {Hall}, J.~R., {Holman}, M.~J., {Koch}, D.~G., {Latham},
  D.~W., {Lopez}, E., {McCauliff}, S., {Miller}, N., {Morehead}, R.~C.,
  {Quintana}, E.~V., {Ragozzine}, D., {Sasselov}, D., {Short}, D.~R., \&
  {Steffen}, J.~H. 2011, \nat, 470, 53

\bibitem[{{Lissauer} {et~al.}(2013){Lissauer}, {Jontof-Hutter}, {Rowe},
  {Fabrycky}, {Lopez}, {Agol}, {Marcy}, {Deck}, {Fischer}, {Fortney}, {Howell},
  {Isaacson}, {Jenkins}, {Kolbl}, {Sasselov}, {Short}, \&
  {Welsh}}]{Lissauer2013}
{Lissauer}, J.~J., {Jontof-Hutter}, D., {Rowe}, J.~F., {Fabrycky}, D.~C.,
  {Lopez}, E.~D., {Agol}, E., {Marcy}, G.~W., {Deck}, K.~M., {Fischer}, D.~A.,
  {Fortney}, J.~J., {Howell}, S.~B., {Isaacson}, H., {Jenkins}, J.~M., {Kolbl},
  R., {Sasselov}, D., {Short}, D.~R., \& {Welsh}, W.~F. 2013, \apj, 770, 131

\bibitem[{{Lithwick} \& {Wu}(2011)}]{Lithwick2011}
{Lithwick}, Y. \& {Wu}, Y. 2011, \apj, 739, 31

\bibitem[{{Lithwick} {et~al.}(2012){Lithwick}, {Xie}, \& {Wu}}]{LXW}
{Lithwick}, Y., {Xie}, J., \& {Wu}, Y. 2012, [LXW] ArXiv

\bibitem[{{Lopez} {et~al.}(2012){Lopez}, {Fortney}, \& {Miller}}]{Lopez}
{Lopez}, E.~D., {Fortney}, J.~J., \& {Miller}, N. 2012, \apj, 761, 59

\bibitem[{{Migaszewski} {et~al.}(2012){Migaszewski}, {S{\l}onina}, \&
  {Go{\'z}dziewski}}]{Migas}
{Migaszewski}, C., {S{\l}onina}, M., \& {Go{\'z}dziewski}, K. 2012, \mnras,
  427, 770

\bibitem[{{Milani} \& {Nobili}(1992)}]{Milani}
{Milani}, A. \& {Nobili}, A.~M. 1992, \nat, 357, 569

\bibitem[{{Milani} {et~al.}(1997){Milani}, {Nobili}, \& {Knezevic}}]{Milani97}
{Milani}, A., {Nobili}, A.~M., \& {Knezevic}, Z. 1997, \icarus, 125, 13

\bibitem[{{Morbidelli} \& {Froeschl{\'e}}(1996)}]{morbi96}
{Morbidelli}, A. \& {Froeschl{\'e}}, C. 1996, Celestial Mechanics and Dynamical
  Astronomy, 63, 227

\bibitem[{{Murray} \& {Dermott}(2000)}]{MD00}
{Murray}, C.~D. \& {Dermott}, S.~F. 2000, {Solar System Dynamics} (Cambridge
  University Press)

\bibitem[{{Murray} \& {Holman}(1997)}]{murray97}
{Murray}, N. \& {Holman}, M. 1997, \aj, 114, 1246

\bibitem[{{Murray} \& {Holman}(1999)}]{MurrayHolman}
---. 1999, Science, 283, 1877

\bibitem[{{Nekhoroshev}(1977)}]{Nek}
{Nekhoroshev}, N.~N. 1977, Russian Mathematical Surveys, 32, 1

\bibitem[{{Ogihara} \& {Kobayashi}(2013)}]{ogihara13}
{Ogihara}, M. \& {Kobayashi}, H. 2013, \apj, 775, 34

\bibitem[{{Owen} \& {Wu}(2013)}]{OwenWu}
{Owen}, J.~E. \& {Wu}, Y. 2013, \apj, 775, 105

\bibitem[{{Papaloizou} \& {Szuszkiewicz}(2005)}]{pap2005}
{Papaloizou}, J.~C.~B. \& {Szuszkiewicz}, E. 2005, \mnras, 363, 153

\bibitem[{{Quillen}(2011)}]{Quillen}
{Quillen}, A.~C. 2011, \mnras, 418, 1043

\bibitem[{{Raymond} {et~al.}(2006){Raymond}, {Quinn}, \& {Lunine}}]{Raymond06}
{Raymond}, S.~N., {Quinn}, T., \& {Lunine}, J.~I. 2006, \icarus, 183, 265

\bibitem[{{Rein}(2012)}]{rein12}
{Rein}, H. 2012, \mnras, 427, L21

\bibitem[{{Smith} \& {Lissauer}(2009)}]{Smith}
{Smith}, A.~W. \& {Lissauer}, J.~J. 2009, \icarus, 201, 381

\bibitem[{{Sussman} \& {Wisdom}(1988)}]{sussman}
{Sussman}, G.~J. \& {Wisdom}, J. 1988, Science, 241, 433

\bibitem[{{Terquem} \& {Papaloizou}(2007)}]{TerquemPap}
{Terquem}, C. \& {Papaloizou}, J.~C.~B. 2007, \apj, 654, 1110

\bibitem[{{Winter} {et~al.}(2010){Winter}, {Mour{\~a}o}, \& {Giuliatti
  Winter}}]{Winter}
{Winter}, O.~C., {Mour{\~a}o}, D.~C., \& {Giuliatti Winter}, S.~M. 2010, \aap,
  523, A67

\bibitem[{{Wisdom} \& {Holman}(1991)}]{Wisdom1991}
{Wisdom}, J. \& {Holman}, M. 1991, \aj, 102, 1528

\end{thebibliography}
\end{document}